\documentclass[aps,twocolumn,prl]{revtex4}
\usepackage{amsmath,amssymb,bm}
\usepackage{graphicx}
\usepackage{epstopdf}
\usepackage{latexsym}

\newcommand{\beq}{\begin{equation}}
\newcommand{\eeq}{\end{equation}}
\newcommand{\beqn}{\begin{eqnarray}}
\newcommand{\eeqn}{\end{eqnarray}}

\begin{document}
\title{Global Phase Diagram for Magnetism and Lattice Distortion of \\ Fe-pnictide Materials}
\author{Yang Qi}
\affiliation{Department of Physics, Harvard University, Cambridge
MA 02138, USA}
\author{Cenke Xu}
\affiliation{Department of Physics, Harvard University, Cambridge
MA 02138, USA}
\date{\today}
\begin{abstract}

We study the global phase diagram of magnetic orders and lattice
structure in the Fe-pnictide materials at zero temperature within
one unified theory, tuned by both doping and pressure. On the low
doping and high pressure side of the phase diagram, there is one
single transition, which is described by a $z=2$ mean field theory
with very weak run-away flows; on the high doping and low pressure
side the transition is expected to split to two transitions, with
one O(3) spin density wave transition followed by a $z = 3$
quantum Ising transition at larger doping. The fluctuation of the
strain field fluctuation of the lattice will not affect the spin
density wave transition, but will likely drive the Ising nematic
order transition more mean field like through a linear coupling,
as observed experimentally in $\mathrm{BaFe_{2-x}Co_xAs_2}$.

%effectively increases the dimension, and hence drives the nematic
%transition more meanfield like,

\end{abstract}

\maketitle

\section{I, Introduction}

The Iron-superconductor, for its potential to shed new light on
the non-BCS type of superconductors, has attracted enormous
interests since early this year. Despite the complexities and
controversies on the superconducting mechanism, the minimal
tight-binding model, or even the exact pairing symmetry of the
cooper pair, these samples do share two common facts: the
tetragonal-orthorhombic lattice distortion and the $(\pi , 0)$
spin density wave (SDW) \cite{La1}. Both effects are suppressed
under doping and pressure, and they seem always to track each
other in the phase diagram. In Ref. \cite{xms2008,kivelson2008},
the lattice distortion is attributed to preformed spatially
anisotropic spin correlation between electrons, without developing
long range SDW $i.e.$ the lattice distortion and SDW both stem
from magnetic interactions. More specifically, the Ising order
parameter $\sigma$ is represented as $\sigma = \vec{\phi}_1 \cdot
\vec{\phi}_2$, $\vec{\phi}_1$ and $\vec{\phi}_2$ are two Neel
orders on the two different sublattices of the square lattice.

Since this order deforms the electron Fermi surface, equivalently,
it can also be interpreted as electronic nematic order.  The
intimate relation between the structure distortion and SDW phase
has gained many supports from recent experiments. It is suggested
by detailed X-ray, neutron and M\"{o}ssbauer spectroscopy studies
that both the lattice distortion transition and the SDW transition
of $\mathrm{LaFeAs(O_{1-x}F_x)}$ are second order
\cite{LaFeAs2ndorder}, where the two transitions occur separately.
However, in undoped $\mathrm{AFe_2As_2}$ with $\mathrm{A = Sr, \
Eu, \ Ba, \ Ca}$, the structure distortion and SDW occur at the
same temperature, and the transition becomes a strong first order
transition \cite{Ba3,Sr1,Sr2,Sr3,Ca2}. Also, recent neutron
scattering measurements on $\mathrm{Fe_{1+y}Se_xTe_{1-x}}$
indicate that in this material the SDW wave vector is $(\pi/2,
\pi/2)$ for both sublattices \cite{dai11} instead of $(\pi, 0)$ as
in 1111 and 122 materials, and the low temperature lattice
structure is monoclinic instead of orthorhombic (choosing one-Fe
unit cell). These results suggest that the SDW and structure
distortion are indeed strongly interacting with each other, and
probably have the same origin. The sensitivity of the location of
the lattice distortion transition close to the quantum critical
point against the external magnetic field (magnetoelastic effect)
can further confirm this unified picture.

\begin{figure}
\includegraphics[width=2.5in]{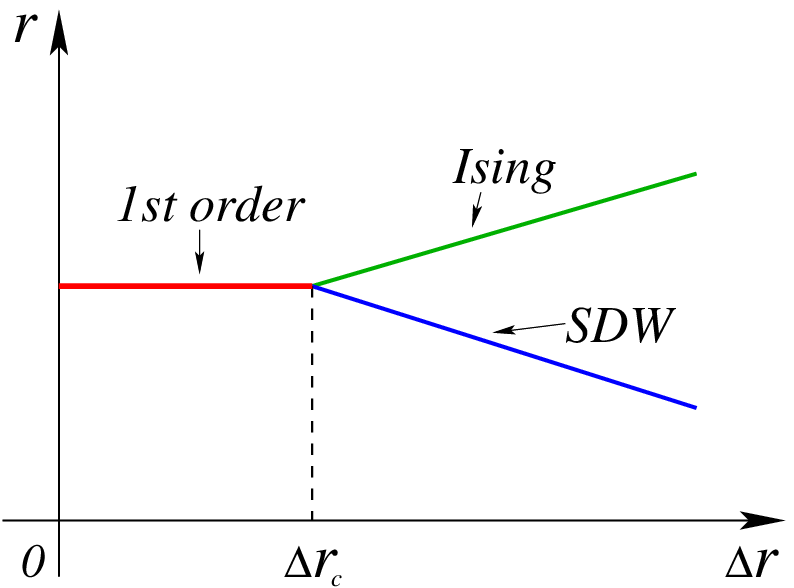}
\caption{The schematic phase diagram of Ginzburg-Landau mean field
theory in Eq. \ref{gl}, plotted against $r = r_\sigma + r_\phi$,
and $\Delta r = r_\phi - r_\sigma$. $r$ is linear with temperature
$T$, while $\Delta r$ is tuned by anisotropy ratio $J_z / J_{in}$.
When $\Delta r$ is small, the interaction between $\vec{\phi}_1$
and $\vec{\phi}_2$ induces a strong first order transition, which
corresponds to the undoped 122 materials with more isotropic
electron kinetics; when $\Delta r$ is large, the transition is
split into two transitions, with an Ising transition followed by
an SDW transition at lower temperature, and this is the case in
the 1111 materials with quasi two dimensional dispersions. The
multicritical point $\Delta r_c$ is determined by $\tilde{u}$.}
\label{pdmf}
\end{figure}

The clear difference between the phase diagrams of 1111 and 122
materials can be naturally understood in the unified theory
proposed in Ref. \cite{xms2008,kivelson2008}. We can write down a
general Ginzburg-Landau mean field theory for $\sigma$,
$\vec{\phi}_1$ and $\phi_2$: \beqn F_{GL} &=& (\nabla_\mu
\sigma)^2 + r_{\sigma}\sigma^2 + \sum_{a = 1}^2
(\nabla_\mu\vec{\phi}_a)^2 + r_{\phi}|\vec{\phi}_a|^2 \cr\cr &+&
\tilde{u}\sigma\vec{\phi}_1\cdot\vec{\phi}_2 + \cdots
\label{gl}\eeqn $r_\sigma$ and $r_\phi$ are tuned by the
temperature. For a purely two dimensional system, the Ising order
which induces the lattice distortion occurs at a temperaure
controlled by the inplane spin coupling $T_{ising} \sim J_{in}$,
while there is no SDW transition at finite temperature; for weakly
coupled two dimensional layers, the Ising transition temperature
is still of the order of the inplane coupling, while the SDW
transition temperature is $T_{sdw}\sim J_{in}/\ln(J_{in}/J_z)$,
with $J_z \ll J_{in}$ representing the interlayer coupling. This
implies that on quasi two dimensional lattices, $T_{ising} \geq
T_{sdw}$, or $\Delta r = r_\phi - r_\sigma$ is large in the
Ginzburg-Landau mean field free energy in Eq. \ref{gl}. In real
systems, The 1111 materials are much more anisotropic compared
with the 122 materials, since the electron band structure
calculated from LDA shows a much weaker $z$ direction dispersion
compared with the 122 samples \cite{122isotropylda}; also the
upper critical field $H_{c2}$ of 122 samples is much more
isotropic \cite{isotropic122}. This justifies treating the 1111
materials as a quasi two dimensional system, while treating the
122 materials as a three dimensional one. When $J_z$ and $J_{in}$
are close enough, $\Delta r$ is small, and the interaction between
the Ising order parameter and the SDW will drive the transition
first order by minimizing the free energy Eq. \ref{gl}. The phase
diagram of free energy Eq. \ref{gl} is shown in Fig. \ref{pdmf}.

Motivated by more and more evidences of quantum critical points in
the Fe-pnictides superconductors \cite{Sm1,Sm2,Ce1,Co1,Co2,Co3},
in this work, we will explore the global phase diagram of magnetic
and nematic orders at zero temperature, tuned by two parameters,
pressure and doping. In section II we will study the phase diagram
for quasi two dimensional lattices, with applications for 1111
materials, and in section III the gear will be switched to the
more isotropic 3d lattices of 122 materials. Section IV will
briefly discuss the effect of the coupling between the Ising
transition and the strain tensor of the lattice, which will drive
the finite temperature Ising nematic transition a mean field
transition, while the SDW transition remains unaffected, as
observed in $\mathrm{BaFe_{2-x}Co_xAs_2}$ \cite{Co1}. The analysis
in our current work are all only based on the symmetry of the
system, and hence independent of the details of the microscopic
model.

\section{II, quasi two dimensional lattice}

\begin{figure}
\includegraphics[width=3.0in]{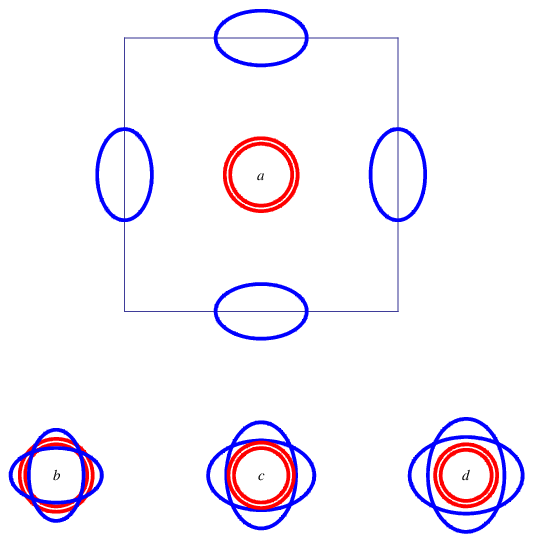}
\caption{$a$, The schematic two dimensional fermi pockets of 1111
materials, the red circles are two very close hole pockets, the
blue ovals are electron pockets. $b - d$, the relative position of
hole and electron pockets after translating by $(\pi, 0)$ and $(0,
\pi)$ in the momentum space, in the low doping, critical doping,
and high doping regimes.} \label{pocketsplot}
\end{figure}

In Ref. \cite{xms2008}, the zero temperature quantum phase
transition was studied for weakly coupled 2d layers with finite
doping. Since the hole pockets and the electron pockets have small
and almost equal size, the slight electron doping would change the
relative size of the electron and hole pockets substantially.
Also, the neutron scattering measurement suggests that the SDW
order wave vector is independent of doping in 1111 materials
\cite{Ce1}. Therefore under doping the low energy particle-hole
pair excitations at wave vector $(\pi, 0)$ are lost very rapidly,
and the spin density wave order parameter at low frequency and
long wavelength limit can no longer decay with particle-hole pairs
(the fermi pockets are schematically showed in Fig.
\ref{pocketsplot}). After integrating out electrons we would
obtain the following $z = 1$ Lagrangian \cite{xms2008}:
%for
%$\vec{\phi}_1$ and $\vec{\phi}_2$
\beqn L &=& \sum_{i = 1}^2\sum_{\mu = \tau, x, y}
\partial_\mu\vec{\phi}_i\cdot \partial_\mu\vec{\phi}_i - r\vec{\phi}_i^2 +
u |\vec{\phi}_i|^4  + L^\prime, \cr\cr L^\prime &=&
\gamma\vec{\phi}_1 (\partial_x^2 - \partial_y^2) \cdot
\vec{\phi}_2 + \gamma_1 |\vec{\phi}_1|^2|\vec{\phi}_2|^2- \alpha
(\vec{\phi}_1\cdot \vec{\phi}_2)^2 , \label{field2} \eeqn which
contains no damping term. The first three terms of the Lagrangian
describe the two copies of 3D O(3) Neel orders on the two
sublattices. The $\alpha$ term is the only relevant term at the 3D
O(3) transition, since it has positive scaling dimension
$\Delta[\alpha] = 0.581$ \cite{vicari2003}. We expect this term to
split the two coinciding O(3) transitions into two transitions, an
O(3) transition and an Ising transition for Ising variable $\sigma
= \vec{\phi}_1 \cdot \vec{\phi}_2$, as observed experimentally in
1111 materials \cite{Ce1}.

The two transitions after splitting are an O(3) transition and an
Ising transition. The O(3) transition belongs to the 3D O(3)
universality class, while the Ising transition is a $z = 3$, $d =
2$ mean field transition. This is because the Ising order
parameter does not double the unit cell, and hence can decay into
particle-hole pairs at momentum $(0, 0)$. The standard
Hertz-Millis theory \cite{hertz1976} would lead to a $z = 3$ mean
field transition \cite{kivelson2001,xms2008}.

Now let us turn on another axis in the phase diagram: the
pressure. Under pressure, the relative size of hole and electron
pockets are not expected to change. Therefore under translation of
$\vec{Q} = (\pi, 0)$ in the momentum space, the hole pocket will
intersect with the electron pocket (Fig. \ref{pocketsplot}$b$),
which leads to overdamping of the order parameters. The decay rate
can be calculated using Fermi's Golden rule: \beqn
\mathrm{Im}[\chi(\omega, q)] &\sim& \int \frac{d^2k}{(2\pi)^2}
[f(\epsilon_{k+q}) - f(\epsilon_{k + \vec{Q}})] \cr\cr &\times&
\delta(\omega - \epsilon_{k+q} + \epsilon_{k+\vec{Q}}) |\langle k
+ Q| \vec{\phi}_{i, q} |k+q \rangle|^2 \ \cr\cr &\sim& \ c_0
\frac{\omega}{|\vec{v}_h \times \vec{v}_e|}. \label{damp} \eeqn
$v_h$ and $v_e$ are the fermi velocity at the points on the hole
and electron pockets which are connected by wave vector $(\pi,
0)$. The standard Hertz-Millis \cite{hertz1976} formalism leads to
a coupled $z=2$ theory in the Euclidean momentum space with
Lagrangian \beqn L_q &=& \sum_{i = 1}^2 \vec{\phi}_i \cdot
(|\omega| + q^2 + r) \vec{\phi}_i  + \gamma\vec{\phi}_1 (q_x^2 -
q_y^2)\cdot \vec{\phi}_2 + L^\prime, \cr\cr L^\prime &=&
 \tilde{A}(|\vec{\phi}_1|^4 + |\vec{\phi}_2|^4) - \alpha
(\vec{\phi}_1\cdot \vec{\phi}_2)^2 + \tilde{C}
|\vec{\phi}_1|^2|\vec{\phi}_2|^2. \label{field3} \eeqn The
parameter $r$ can be tuned by the pressure. The Ising symmetry of
$\sigma = \vec{\phi}_1 \cdot \vec{\phi}_2$ on this system
corresponds to transformation \beqn x &\rightarrow& y, \ \ y
\rightarrow x, \cr\cr \vec{\phi}_1 &\rightarrow& \vec{\phi}_1, \ \
\vec{\phi}_2 \rightarrow - \vec{\phi}_2, \ \ \sigma \rightarrow -
\sigma. \label{symm12}\eeqn This Ising symmetry forbids the
existence of term $\vec{\phi}_1 \cdot \vec{\phi}_2$ in the
Lagrangian, while the mixing term $\gamma\vec{\phi}_1 (q_x^2 -
q_y^2)\cdot \vec{\phi}_2$ is allowed.

\begin{figure}
\includegraphics[width=2.5in]{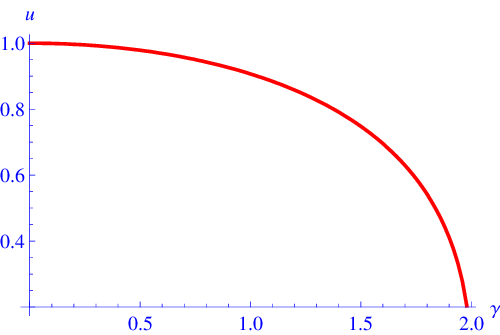}
\caption{The plot of $u$ in Eq. \ref{rg} against anisotropic
dispersion coefficient $\gamma$, between the isotropic limit
$\gamma = 0$ to anisotropic dispersion with $\gamma = 1.95$}
\label{ufunction}
\end{figure}

We can diagonalize the quadratic part of this Lagrangian by
defining $\vec{\phi}_A = (\vec{\phi}_1 + \vec{\phi}_2)/\sqrt{2}$
and $\vec{\phi}_B = (\vec{\phi}_1 - \vec{\phi}_2)/\sqrt{2}$: \beqn
L_q &=& \vec{\phi}_A \cdot (|\omega| + (1 - \frac{\gamma}{2})q^2_x
+ (1+\frac{\gamma}{2})q_y^2 + r) \vec{\phi}_A  \cr\cr &+&
\vec{\phi}_B \cdot (|\omega| + (1+\frac{\gamma}{2}) q^2_x + (1 -
\frac{\gamma}{2})q_y^2 + r) \vec{\phi}_B  + L^\prime, \cr\cr
L^\prime &=&
 A(|\vec{\phi}_A|^4 + |\vec{\phi}_B|^4) +B
(\vec{\phi}_A\cdot \vec{\phi}_B)^2 + C
|\vec{\phi}_A|^2|\vec{\phi}_B|^2. \label{field21}\eeqn After the
redefinition, the Ising transformation becomes \beqn  x
&\rightarrow& y, \ \ y \rightarrow x, \cr\cr \vec{\phi}_A
&\rightarrow& \vec{\phi}_B, \ \ \vec{\phi}_B \rightarrow -
\vec{\phi}_A, \ \ \sigma \rightarrow - \sigma. \label{symmAB}\eeqn
Naively all three quartic terms $A$, $B$ and $C$ are marginal
perturbations on the $z = 2$ mean field theory, a coupled
renormalization group (RG) equation is required to determine the
ultimate fate of these terms. Notice that the anisotropy of the
dispersion of $\vec{\phi}_A$ and $\vec{\phi}_B$ cannot be
eliminated by redefining space and time, therefore the number
$\gamma$ will enter the RG equation as a coefficient. The final
coupled RG equation at the quadratic order for $A$, $B$ and $C$
reads: \beqn \frac{dA}{d\ln l} &=&  - 22A^2 - \frac{1}{2}B^2 -
\frac{3}{2} C^2 - BC, \cr\cr \frac{dB}{d\ln l} &=&  - 5 u B^2 - 8
AB - 8u BC, \cr\cr \frac{dC}{d\ln l} &=&  - u B^2 - 4 AB - 20AC -
4u C^2. \label{rg} \eeqn $u$ is a smooth function of $\gamma$,
which decreases smoothly from $u = 1$ in the isotropic limit with
$\gamma = 0$ to $u = 0$ in the anisotropic limit with $\gamma = 2$
(Fig .\ref{ufunction}). The self-energy correction from the
quartic terms will lead to the flow of the anisotropy ratio
$\gamma$ under RG, but the correction of this flow to the RG
equation Eq. \ref{rg} is at even higher order.

\begin{figure}
\includegraphics[width=2.8in]{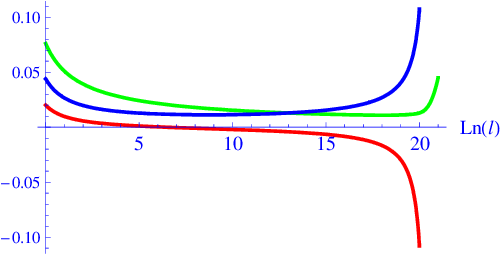}
\caption{The solution of the RG equation Eq. \ref{rg}. All three
quartic perturbations decrease first, then increase and finally
become nonperturbative. The run-away flow is weaker than
marginally relevant perturbations.} \label{rgplot}
\end{figure}

The typical solution of the RG equation Eq. \ref{rg} is plotted in
Fig. \ref{rgplot}. One can see that the three parameters $A$, $B$
and $C$ all have run-away flows and eventually become
nonperturbative, and likely drive the transition weakly first
order. However, the three coefficients will first decrease and
then increase under RG flow. This behavior implies that this
run-away flow is extremely weak, or more precisely even weaker
than marginally relevant perturbations, because marginally
relevant operators will still monotonically increase under RG
flow, although increases slowly. Therefore in order to see this
run-away flow, the correlation length has to be extremely long
$i.e.$ the system has to be very close to the transition, so the
transition remains one single second order mean field transition
for a very large length and energy range. At the finite
temperature quantum critical regime, the standard scaling
arguments lead to the following scaling laws of physical
quantities like specific heat and the spin lattice relaxation rate
of NMR contributed by the quantum critical modes \cite{si2003}:
\beqn C_v \sim T\ln(\frac{1}{T}), \ \ \ \frac{1}{T_1} \sim
\mathrm{Const}. \eeqn These scaling behaviors are obtained from
ignoring the quartic perturbations. The quartic terms are marginal
for a rather large energy scale (Fig. \ref{rgplot}), therefore to
precisely calculate the physical quantities one should perform a
perturbation theory with constant $A$, $B$ and $C$, which may lead
to further logarithmic corrections to the scaling laws.

The $\hat{z}$ direction tunnelling of $\vec{\phi}_A$ and
$\vec{\phi}_B$ between layers has so far been ignored, which is
also a relevant perturbation at the $z = 2$ mean field fixed
point. The $\hat{z}$ direction tunnelling is written as $J_z
\vec{\phi}_{a, z} \cdot \vec{\phi}_{a, z+1}$, which has scaling
dimension 2 at the $z = 2$, $d = 2$ mean field fixed point, and it
becomes nonperturbative when \beqn \frac{J_{in}}{J_z} \sim
(\frac{\xi}{a})^2 \sim r^{-1}. \eeqn This equation implies that if
the tuning parameter $r$ is in the small window $r < J_z /
J_{in}$, the transition crossover back to a $z = 2$, $d = 3$
transition, where all the quartic perturbation $A$, $B$ and $C$
are irrelevant. Since at the two dimensional theory these quartic
terms are only weakly relevant up to very long length scale, in
the end the interlayer coupling $J_z$ may win the race of the RG
flow, and this transition becomes one stable mean field second
order transition.

\begin{figure}
\includegraphics[width=2.7in]{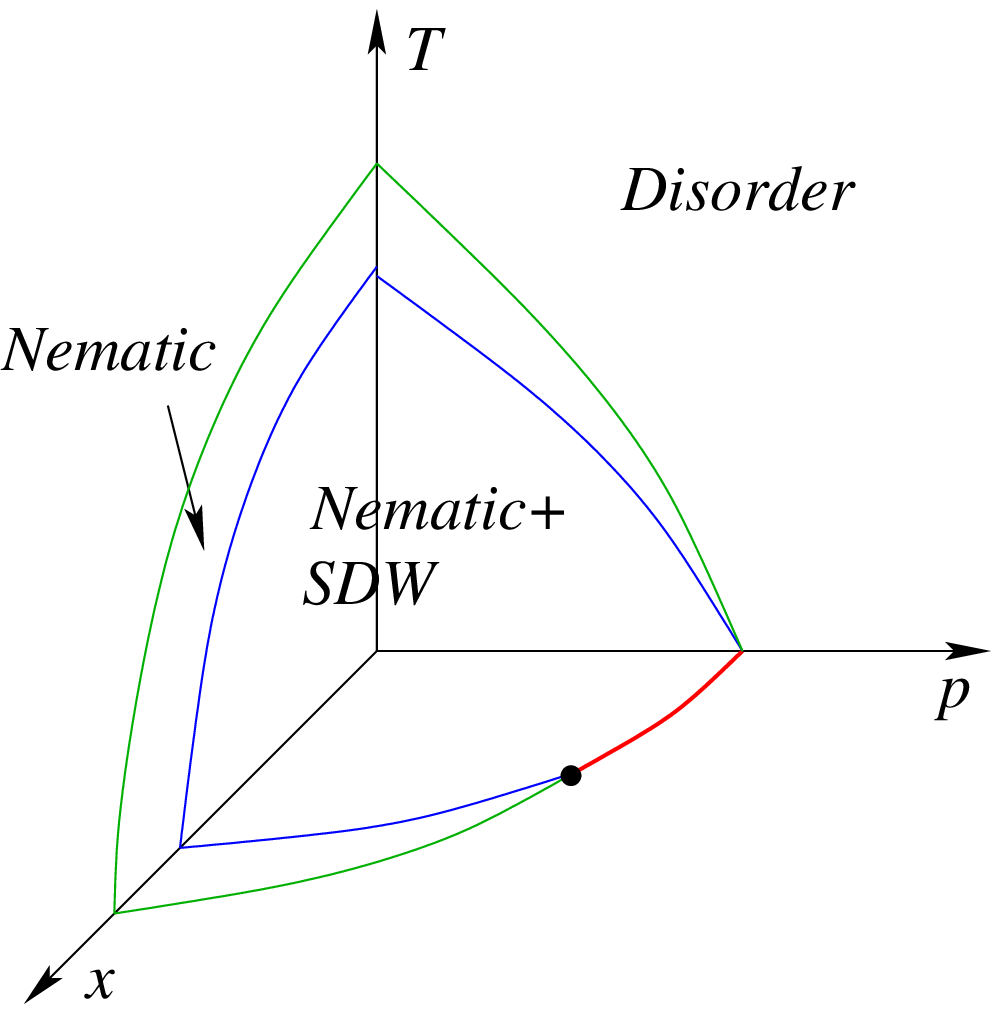}
\caption{The global phase diagram of quasi two dimensional
materials, with applications for 1111 materials. The finite
temperature transition is always split to an Ising nematic
transition and a SDW transition. The zero temperature transitions
depend on the doping and pressure. In the high doping and low
pressure side, the transition is split to two, as observed in
experiments; in the low doping and high pressure side, there is
one single transition very close to the mean field solution. A
multicritical point where the three transition lines merge is
identified, which is expected to be a strongly coupled fixed
point.} \label{pdgm}
\end{figure}

Now we have a global two dimensional phase diagram whose two axes
are doping and pressure. The two second order transition lines in
the large doping and low pressure side will merge to one single
mean field transition line in the low doping and high pressure
side of the phase diagram. Then inevitably there is a
multicritical point where three lines merge together. At this
multicritical point, the hole pockets will just touch the electron
pocket after translating by wave vector $(\pi, 0)$ (Fig.
\ref{pocketsplot}$c$). Now the SDW order parameter $\vec{\phi}_A$
and $\vec{\phi}_B$ can still decay into particle-hole pairs, the
Fermi's Golden rule and the lattice symmetry lead to the following
overdamping term in the Lagrangian: \beqn L_q &=&
(\frac{|\omega|}{\sqrt{|q_x|}} + g\frac{|\omega|}{\sqrt{|q_y|}} )
|\vec{\phi}_A|^2 \cr\cr &+& (g \frac{|\omega|}{\sqrt{|q_x|}} +
\frac{|\omega|}{\sqrt{|q_y|}} ) |\vec{\phi}_B|^2 + \cdots \eeqn
$g$ is a constant, which is in general not unity because the
system only enjoys the symmetry Eq. \ref{symmAB}. The naive
power-counting shows that this field theory has dynamical exponent
$z = 5/2$, which makes all the quartic terms irrelevant. However,
since the hole pockets and electron pockets are tangential after
translating $(\pi, 0)$, the expansion of the mean field free
energy in terms of the order parameter $\vec{\phi}_A$ and
$\vec{\phi}_B$ contains a singular term $L_s \sim
|\vec{\phi}_A|^{5/2} + |\vec{\phi}_B|^{5/2}$, which becomes very
relevant at this naive $z = 5/2$ fixed point. Similar singular
term was found in the context of electronic nematic-smectic
transition \cite{fradkinsmectic}. The existence of this singular
term implies that, it is inadequate to start with a pure Bose
theory by integrating out fermions, one should start with the
Bose-Fermi mixed theory, with which perform the RG calculation. We
will leave this sophisticated RG calculation to the future work,
right now we assume this multicritical point is a special strongly
interacting fixed point. The schematic three dimensional global
phase diagram is shown in Fig. \ref{pdgm}.

\begin{figure}
\centering
\includegraphics[width=2.9in]{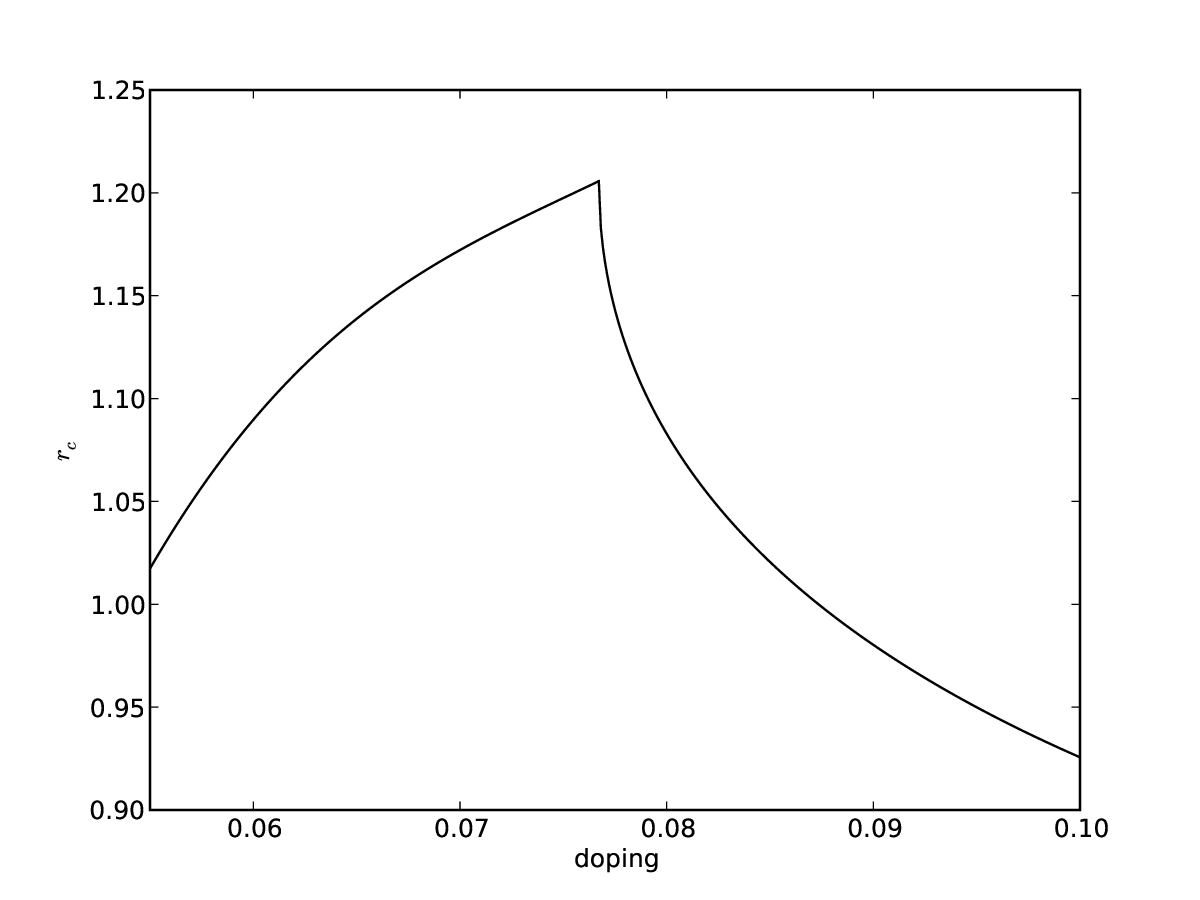}
\caption{Numerical results of $r_c$ of $\vec{\phi}_a$ due to
coupling to electrons. $x$ axis is the electron doping. The peak
of this curve corresponds to the critical doping $x_c =7.6\%$
where electron and hole pockets touch each other after translating
the hole pockets by the SDW wave vector. The two pockets intersect
(separate) if doping is smaller (larger) than this critical
doping. If $x > x_c$, the transition is split into two transitions
by quantum fluctuations; if $x < x_c$, the transition is a $z =
2$, $d = 2$ transition with a very weak run-away flow in 2d.}
\label{fig:rplot}
\end{figure}

\begin{figure}
\centering
\includegraphics[width=2.8in]{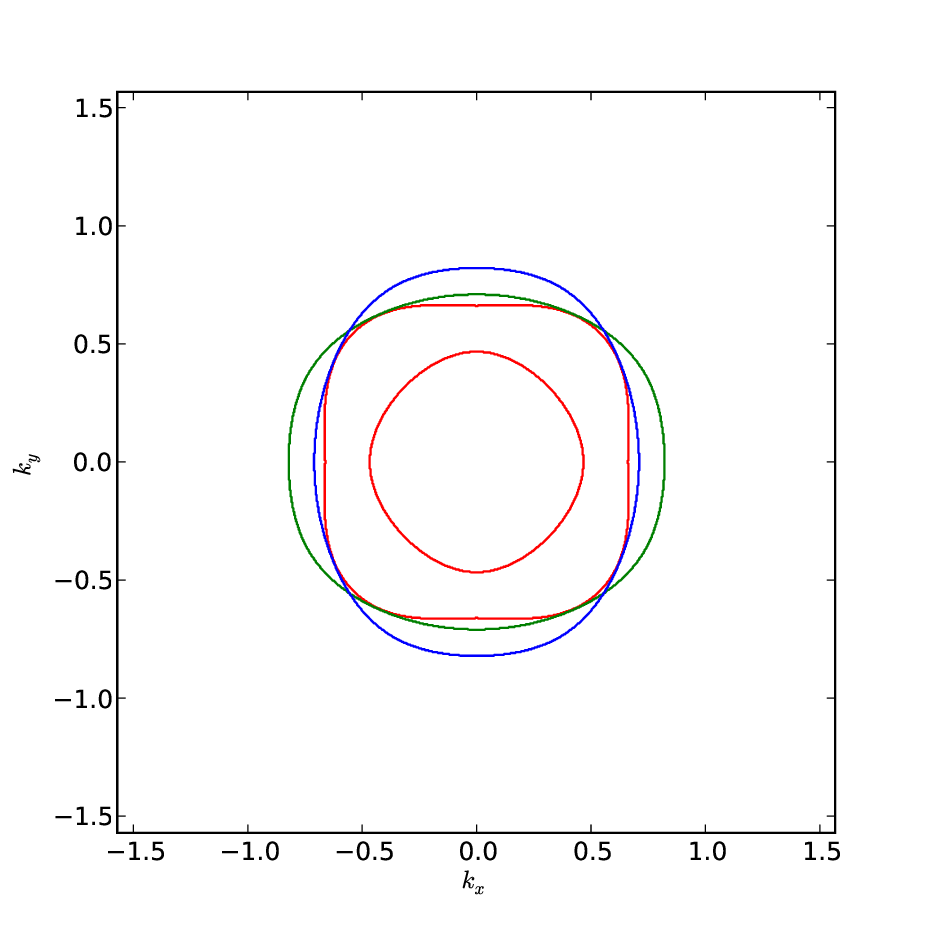}
\caption{The plot of the hole and electron pockets after
translating the hole pockets by the SDW wave vector, at the
critical doping $x = 7.6\%$. The green circle is electron pocket
located around $(0, \pi)$ and the blue one is electron pocket
located around $(\pi, 0)$.} \label{fig:crtfs}
\end{figure}

In real system, due to the more complicated shape of the electron
and hole pockets, with increasing doping the pockets will
experience cutting and touching several times after translating by
$(\pi, 0)$ in the momentum space. We have used a five-band model
developed in Ref. \cite{kurokimodel} with all the $d-$orbitals on
the Fe atoms, and calculated the mean field phase diagram close to
the critical doping. The order parameter $\vec{\phi}_a$ couple to
the electrons at the Fermi surface as: $\sum_k \vec{\phi}_a \cdot
c^\dagger_{k}\vec{\sigma}c_{k+\vec{Q}} + H.c.$. The mean field
energy of electrons due to nonzero spin order parameter
$\vec{\phi}_a$ will renormalize $r$ in field theory Eq.
\ref{field21}, and hence the critical $r_c$ depends on the shape
of the Fermi surface, which is tuned by doping. The critical $r_c$
is expected to be proportional to the critical pressure $p_c$ in
the global phase diagram. $r_c$ as a function of doping is plotted
in Fig. \ref{fig:rplot}, and the shapes of the Fermi pockets at
the critical doping $x = 7.6\%$ is plotted in Fig.
\ref{fig:crtfs}.

The $z = 2$ quantum critical behavior discussed in this section is
only applicable to small enough energy scale. First of all, the
damping term $|\omega|$ always competes with a quadratic term
$\omega^2$ in the Lagrangian, and at small enough energy scale the
linear term dominates. If we assume the coupling between the spin
order parameter $\vec{\phi}_a$ and the electrons is of the same
order as the effective spin interaction $J$, the damping rate is
linear with $\sim J^2 \omega/E_f^2$, while the quadratic term is $
\sim \omega^2 / J$. Therefore the frequency should be smaller than
$J^3/E_f^2$ in order to apply the $z = 2$ field theory Eq.
\ref{field21}. The value of $J$ has been calculated by LDA
\cite{xiang2008}, and also measured with inelastic neutron
scattering \cite{inelasticneutron}, and both approaches indicate
that $J \sim 50$meV. $E_f$ is the Fermi energy of the Fermi
pockets, which is of the order of $200$meV. Therefore the
frequency-linear damping term will dominate the the frequency
quadratic term in the Lagrangian as long as $\omega < 3$meV.

The damping rate of order parameters $\vec{\phi}_a$ is calculated
assuming the Fermi surface can be linearly expanded close to the
intersection point after translation in the momentum space, the
criterion to apply this assumption depends on the details on the
Fermi surface. In the particular situation under discussion, this
crossover energy scale is $\omega \sim 30$meV in the undoped
material, which is larger than the upper limit of 3meV we obtained
previously. Therefore the ultraviolet cut-off of field theory Eq.
\ref{field21} is estimated to be $3$meV.

\section{III, three dimensional lattices}

As mentioned in the introduction, compared with the 1111
materials, the 122 materials are much more isotropic, so we will
treat this family of materials as a three dimensional problem. If
after translation by $(\pi, 0)$ the hole pockets intersect with
the electron pockets, the zero temperature quantum transition is
described a $z = 2$, $d = 3$ transition with analogous Lagrangian
as Eq. \ref{field3}, which becomes a stable mean field transition.
The finite temperature transition is described by two copies of
coupled 3D O(3) transition. If the finite temperature transition
is split into two transitions close to the quantum critical point,
as observed in $\mathrm{BaFe_{2-x}Co_xAs_2}$ \cite{Co1}, one can
estimate the size of the splitting close to the quantum critical
point. These two transitions, as explained before, is driven by
the only relevant perturbation $\alpha (\vec{\phi}_1 \cdot
\vec{\phi}_2)^2$ at the coupled 3D O(3) transition, because the
Ising order parameter is obtained by minimizing this term through
Hubbard-Stratonovich transformation. The scaling dimension of
$\alpha$ at the 3D O(3) transition is $\Delta[\alpha] = 0.581$,
while $\alpha$ at the $z = 2$, $d = 3$ mean field fixed point has
dimension $-1$. Therefore close to the quantum critical regime, to
estimate the effect of $\alpha$ one should use the renormalized
value $\alpha_R \sim \alpha \xi^{-1} \sim \alpha r^{1/2}$. The
size of the splitting of the finite temperature transition close
to the quantum critical point can be estimated as \beqn
\frac{\Delta T_c}{T_c} \sim \alpha_R^{1/(\nu\Delta[\alpha])} \sim
\alpha^{1/(\nu\Delta[\alpha])} r^{1/(2\nu\Delta[\alpha])}. \eeqn
$\nu$ is the exponent defined as $\xi \sim t^{-\nu}$ at the 3D
O(3) universality class. $T_c$ still scales with $r$ in terms of a
universal law $T_c \sim r^{z/(d - 2 +z)}$. The number $\alpha$ can
be estimated in a $J_1 - J_2$ Heisenberg model on the square
lattice as introduced in Ref. \cite{qij1j2}, the value is given by
$\alpha \sim J_1^2/J_2^2$. However, $J_1 - J_2$ model is not
designed for describing a metallic phase, so the legitimacy of
applying the $J_1-J_2$ model to Fe-pnictides is still under
debate. In the finite temperature quantum critical regime, the
specific heat, NMR relaxation rate $1/T_1$ scale as \beqn C_v \sim
T^{3/2}, \ \ \ \frac{1}{T_1} \sim T^{1/2}. \eeqn

The similar analysis also applies when the finite temperature
transition is one single first order transition, which is the more
common situation in 122 materials. One can estimate the jump of
the lattice constant, and the jump of the SDW order parameter at
the finite temperature first order transition close to the quantum
critical point as \beqn \delta \vec{\phi}_{sdw} &\sim&
\alpha^{\beta/(\nu\Delta[\alpha])} r^{\beta/(2\nu\Delta[\alpha])},
\cr \cr \delta a &\sim& \alpha^{2\beta/(\nu\Delta[\alpha])}
r^{\beta/(\nu\Delta[\alpha])}. \eeqn $a$ is the lattice constant,
which is linear with the Ising order parameter $\vec{\phi}_1 \cdot
\vec{\phi}_2$. $\beta$ is the critical exponent at the 3D O(3)
transition defined as $ \langle\vec{\phi}_{sdw}\rangle \sim
t^{\beta}$.

If under doping the hole pockets and electron pockets do not
intersect (which depends on the details of $\hat{z}$ direction
dispersions), this transition becomes two copies of coupled $z =
1$, $d = 3$ transition with three quartic terms $A$, $B$ and $C$:
\beqn L_q &=& \vec{\phi}_A \cdot (\omega^2 + (1 -
\frac{\gamma}{2})q^2_x + (1+\frac{\gamma}{2})q_y^2 + q_z^2 + r)
\vec{\phi}_A  \cr\cr &+& \vec{\phi}_B \cdot (\omega^2 +
(1+\frac{\gamma}{2}) q^2_x + (1 - \frac{\gamma}{2})q_y^2 + q_z^2 +
r) \vec{\phi}_B  \cr\cr &+& L^\prime, \cr\cr L^\prime &=&
 A(|\vec{\phi}_A|^4 + |\vec{\phi}_B|^4) +B
(\vec{\phi}_A\cdot \vec{\phi}_B)^2 + C
|\vec{\phi}_A|^2|\vec{\phi}_B|^2. \label{field4}\eeqn The coupled
RG equation of the quartic terms are exactly the same as the one
in Eq. \ref{rg}, therefore this free energy is also subjected to
an extremely weak run-away flow, which is negligible unless the
length scale is large enough. Again, one can estimate the
universal scaling behavior in the quantum critical regime
contributed by the quantum critical modes: \beqn C_v \sim T^{3}, \
\ \ \frac{1}{T_1} \sim T. \eeqn

\section{IV, coupling to a soft lattice}

Recent specific heat measurement on $\mathrm{BaFe_{2-x}Co_xAs_2}$
reveals two close but separate transitions at finite temperature,
with a sharp peak at the SDW transition, and a discontinuity at
the lattice distortion transition \cite{Co1}. A discontinuity of
specific heat is a signature of mean field transition, in contrast
to the sharp peak of Wilson-Fisher fixed point in 3 dimensional
space. The specific heat data suggest that the nature of the Ising
nematic transition is strongly modified from the Wilson-Fisher
fixed point, while SDW transition is unaffected. In the following
we will attribute this difference to the lattice strain field
fluctuations.

%which can be naturally explained with the analysis of the
%elasticity of the lattice above. The sharp peak at the SDW
%transition implies that the lattice is much less important at the
%3D O(3) transition, which is again consistent with our analysis
%above.

The SDW transition at finite temperature should belong to the 3D
O(3) transition ignoring the lattice. The O(3) order parameter
$\vec{\phi}$ couples to the the lattice strain field with a
quadratic term \cite{dh}: \beqn |\vec{\phi}|^2(\partial_x u_x +
\partial_y u_y + \lambda^\prime \partial_z u_z), \eeqn which after integrating out the
displacement vector generates a singular long range interaction
between $|\vec{\phi}|^2$ in the real space: \beqn \int d^3r
d^3r^\prime g|\vec{\phi}|^2_r \frac{f(\vec{r} -
\vec{r}^\prime)}{|r - r^\prime|^3}|\vec{\phi}|^2_{r^\prime}. \eeqn
$f$ is a dimensionless function which depends on the direction of
$\vec{r} - \vec{r}^\prime$. The scaling dimension of $g$ is
$\Delta[g] = 2/\nu - 3$, and $\nu$ is the standard exponent at the
3D O(3) transition, which is greater than $2/3$ according to
various types of numerical computations \cite{vicari2003}.
Therefore this long range interaction is irrelevant at the 3D O(3)
transition, and by coupling to the strain field of the lattice,
the SDW transition is unaffected. However, if the SDW has an Ising
uniaxial anisotropy, the SDW transition becomes a 3D Ising
transition with $\nu < 2/3$, and the strain field would lead to a
relevant long range interaction.

However, since the symmetry of the Ising order parameter $\sigma$
is the same as the shear strain of the lattice, the strain tensor
will couple to the coarse grained Ising field $\Phi$ as \beqn
F_{\Phi, \vec{u}} &=& \tilde{\lambda} \Phi (\partial_x u_y +
\partial_y u_x) + \cdots \eeqn $\vec{u}$ is the displacement
vector, The ellipses are all the elastic modulus terms. Notice
that we have rotated the coordinates by 45 degree, since the true
unit cell of the system is a two Iron unit cell. After integrating
out the displacement vector $\vec{u}$, the effective free energy
of $\Phi$ gains a new singular term at small momentum: \beqn
F_{\theta, \phi} \sim
 f(\theta, \phi)|\Phi_k|^2 . \eeqn $f$ is a function of spherical
 coordinates
 $\theta$ and $\phi$ defined as $(k_x, k_y, k_z) = k (\cos(\theta)\cos(\phi),
\cos(\theta)\sin(\phi), \sin(\theta))$, but $f$ is independent of
the magnitude of momentum $\vec{k}$. By tuning the uniform
susceptibility $r$, at some spherical angle of the space the
minima of $f$ start to condense, we will call these minima as
nodal points. These nodal points are isolated from each other on
the two dimension unit sphere labelled by the solid angles
$\theta, \ \phi$, and are distributed symmetrically on the unit
sphere $(\theta, \phi)$ according to the lattice symmetry
transformation. Now suppose one nodal point of $f$ is located at
$(\theta_0, \phi_0)$, we rotate the $\hat{z}$ direction along
$(\theta_0, \phi_0)$, and expand $f$ at this nodal point in terms
of $\tilde{\theta} = \theta - \theta_0$, the whole free energy can
be written as \beqn F = \int q^2 dq \tilde{\theta} d\tilde{\theta}
( q^2 + \lambda\tilde{\theta}^2 + r)|\Phi_{q, \tilde{\theta}}|^2 +
O(\Phi^4). \label{5d}\eeqn Notice that if $f(\theta_0, \phi_0)$ is
a nodal point, then $f(\pi - \theta_0, \pi + \phi_0)$ has to be
another nodal point. The naive power counting shows that
effectively the spatial dimension of this field theory Eq.
\ref{5d} is $D=5$, and the scaling dimension of $\Phi_{q, \theta}$
is $-7/2$. The quartic term $\Phi^4$ takes an unusual form in the
new momentum space of $\ q, \ \tilde{\theta}$, but the
straightforward power counting indicates that it is still an
irrelevant operator. Therefore the strain tensor fluctuation
effectively increases the dimension by two, which drives the
transition a mean field transition.

Another way to formulating this effective 5 dimensional theory is
that, close to the minimum $\theta = 0$, if the scaling dimension
of $k_z$ is fixed to be 1, then $k_x, \ k_y \sim k \theta$
effectively have scaling dimension 2. Therefore expanded at the
minimum the quadratic part of the free energy of $ \Phi$ reads:
\beqn F = \int dk_x dk_y dk_z (\frac{k_x^2 + k_y^2}{k_z^2} + k_z^2
+ r)|\Phi_k|^2 + \cdots \eeqn The total dimension is still 5,
considering $\Delta[k_x] = \Delta[k_y] = 2\Delta[k_z] = 2$. All
the other momentum dependent terms in the free energy are
irrelevant.

The symmetry of the lattice allows multiple degenerate nodal
points of function $f(\theta, \phi)$ on the unit sphere labelled
by solid angles. If the only nodal points are north and south
poles $\theta = 0, \ \pi$, which is allowed by the tetragonal
symmetry of the lattice, the theory becomes a precise 5
dimensional theory. However, the symmetry of the system also
allows four stable degenerate nodal points on the equator, for
instance at $(\pi/2, n \pi/2)$ with $n = 0 \sim 3$. Close to nodal
points $n = 0, \ 2$, $\Delta[k_z] = \Delta[k_y] = 2\Delta[k_x] =
2$, while close to nodal points $n = 1, \ 3$, $\Delta[k_z] =
\Delta[k_x] = 2\Delta[k_y] = 2$. Therefore the scattering between
these nodal points complicates the naive counting of the scaling
dimensions, although $\Delta[k_z] = 2$ is still valid. The
transition in this case may still be a stable mean field
transition, but more careful analysis of the loop diagrams is
demanded to be certain. Let us denote the $\Phi$ mode at $(\pi/2,
0)$ and $(\pi/2, \pi)$ as $\Phi_1$ and $\Phi_1^\ast$, and denote
$(\pi/2, \pi/2)$ and $(\pi/2, -\pi/2)$ modes as $\Phi_2$ and
$\Phi_2^\ast$, the expanded free energy reads \beqn F &=& \int
dk_x dk_y dk_z (\frac{k_z^2 + k_y^2}{k_x^2} + k_x^2 + r)|\Phi_{1,
k}|^2 \cr\cr &+&  (\frac{k_z^2 + k_x^2}{k_y^2} + k_y^2 +
r)|\Phi_{2, k}|^2 \cr\cr &+& \sum_{a = 1}^2 \delta(\sum_{i=1}^4
{\vec{k}}_i)g\Phi_{a,k_1}\Phi_{a,k_2}\Phi_{a,k_3}\Phi_{a,k_4}
\cr\cr &+& \delta(\sum_{i=1}^4
{\vec{k}}_i)g_1\Phi_{1,k_1}\Phi_{1,k_2}\Phi_{2,k_3}\Phi_{2,k_4}
\cr\cr &+& \delta(\sum_{i=1}^4
{\vec{k}}_i)g_2\Phi_{1,k_1}\Phi_{2,k_2}\Phi_{2,k_3}\Phi_{2,k_4}
\cr\cr &+& \delta(\sum_{i=1}^4
{\vec{k}}_i)g_2\Phi_{1,k_1}\Phi_{1,k_2}\Phi_{1,k_3}\Phi_{2,k_4}.
\eeqn The $g_1$ and $g_2$ terms describe the scattering between
different nodal points. To see whether the mean field transition
is stable, one can calculate the one-loop corrections to $g_1$ and
$g_2$, and none of the loops introduces nonperturbative divergence
in the infrared limit. This analysis suggests that the Ising
nematic transition is a mean field transition even with multiple
nodal points of function $f(\theta, \phi)$ on the equator.

%$\Phi$ will linearly couple to the shear strain of the lattice,
%while $\Phi^2$ will couple to the bulk strain $\partial_x u_x +
%\partial_y u_y$ \cite{dh}. By integrating out the displacement field
%$\vec{u}$, other than a regular $\Phi^4$ term, a singular term
%$\Phi^2_{q,\theta}\Phi^2_{q,-\theta}[\cos(2\theta)]^2$ is
%generated. If we expand this term at the nodal point $\theta =
%\pi/4$, this term behaves as if a term with second derivative:
%$\Phi^2_{q,\theta}\Phi^2_{q,-\theta}(\delta\theta)^2$, which
%becomes clearly irrelevant at the 4D mean field fixed point.
%Therefore we conclude that the two dimensional strain field
%fluctuation will lead to a 4D mean field second order transition
%for the Ising nematic order parameter.

\section{V, summaries and extensions}

In this work we studied the global phase diagram of the magnetic
order and lattice distortion of the Fe-pnictides superconductors.
A two dimensional and three dimensional formalisms were used for
1111 and 122 materials respectively. The superconductivity was
ignored so far in this material. If the quantum critical points
discussed in this paper occur inside the superconducting phase,
our results can be applied to the case when superconducting phase
is suppressed. For instance, in 1111 materials, if a transverse
magnetic field higher than $H_{c2,ab}$ is turned on, the field
theory Eq. \ref{field2} and Eq. \ref{field21} become applicable.
If the $T_c$ of the superconductor is lower than the ultraviolet
cut-off of our field theory, the scaling behavior predicted in our
work can be applied to the temperature between $T_c$ and the
cut-off. Inside the superconducting phase, the nature of the
transition may be changed. In 122 materials, the ARPES
measurements on single crystals indicate that the fermi pockets
are fully gapped in the superconducting phase \cite{swave122},
therefore the magnetic and nematic transitions are described by
the $z = 1$, $d = 3$ field theory Eq. \ref{field4}, which is an
extremely weak first order transition. In 1111 materials, although
many experimental facts support a fully gapped fermi surface,
$d-$wave pairing with nodal points is still favored by the Andreev
reflection measurements \cite{andreevSm,STMSm}. The nematic
transition is the background of $d-$wave superconductor is studied
in Ref. \cite{kimnematic,huh2008,xunematic}.

In most recently discovered 11 materials
$\mathrm{Fe_{1+y}Se_xTe_{1-x}}$, the SDW and lattice distortion
are both different from the 1111 and 122 materials \cite{dai11}.
The SDW state breaks the reflection symmetries about both $x = y$
line and $\hat{x}$ axis $i.e.$ there are two different Ising
symmetries broken in the SDW state, the ground state manifold is
$S^2 \times Z_2 \times Z_2$. In this case, the classical and
quantum phase diagrams are more interesting and richer, and since
the order moments of the SDW in 11 materials are much larger than
1111 and 122 materials (about $2\mu_B$), a lattice Heisenberg
model with nearest neighbor, 2nd nearest neighbor, and 3rd nearest
neighbor interactions ($J_1 - J_2 - J_3$) may be adequate in
describing 11 materials, as was studied in Ref. \cite{andrei2008}.

Besides the quantum phase transitions studied in our current work,
a quantum critical point is conjectured between the
$\mathrm{P-}$based and $\mathrm{As-}$based materials
\cite{sicritical}, the field theory of this quantum critical point
is analogous to Eq. \ref{field3}. The formalism used in our work
is also applicable to phase transitions in other strongly
correlated materials, for instance the spin-dimer material
$\mathrm{BaCuSi_2O_6}$, which under strong magnetic field develops
long range XY order interpreted as condensation of spin triplet
component $S^z = -1$ \cite{BaCuSiO}. This quantum critical point
also has dynamical exponent $z = 2$, although the frequency linear
term is from the Larmor precession induced by the magnetic field,
instead of damping with particle-hole excitations. The frustration
between the nearest neighbor layers in this material introduces an
extra Ising symmetry between the even and odd layers besides the
XY spin symmetry, therefore the quartic terms of this field theory
are identical with Eq. \ref{field21}. The RG equations of these
quartic terms are much simpler than Eq. \ref{rg}, because only the
``ladder" like Feynman diagrams need to be taken into account
\cite{sachdevbook}. We will study the material
$\mathrm{BaCuSi_2O_6}$ in detail in a future work
\cite{xuBaCuSiO}.

%If we isolate two nearest neighbor layers from this material with
%XY order parameter $\vec{\phi}_1$ and $\vec{\phi}_2$, The RG
%equations of these quartic terms are much simpler than Eq.
%\ref{rg}, because only the ``ladder" like Feynman diagrams should
%be taken into account \cite{sachdevbook}. In this case some of the
%quartic terms will also first decrease and then increase under RG
%flow, and hence are extremely weak relevant perturbations. This is
%consistent with the experimental observations that the transition
%remains a perfect $z = 2$, $d = 2$ mean field transition until
%untra-low temperature \cite{BaCuSiO}.

\begin{acknowledgments}

We thank Subir Sachdev and Qimiao Si for heplful discussions. We
especially appreciate Bert Halperin for educating us about his
early work on phase transitions on soft cubic lattice \cite{dh}.

\end{acknowledgments}

\end{document}